\documentclass[twocolumn]{article}
\usepackage{mlspconf}

\usepackage{amsmath,amssymb}
\usepackage{rotating}
\usepackage{graphicx}
\usepackage[bookmarks=false]{hyperref}

\toappear{2016 IEEE International Workshop on Machine Learning for Signal Processing, Sept.\ 13--16, 2016, Salerno, Italy}

\def\longversion{1} 

\title{Bird detection in audio: a survey and a challenge}

\name{
\begin{tabular}{p{4.5cm}p{4.5cm}p{4.1cm}p{4.1cm}}
\centering Dan Stowell\sthanks{Supported by EPSRC fellowship EP/L020505/1.} &
\centering Mike Wood\sthanks{Supported by NERC grant NE/L000520/1.} &
\centering Yannis Stylianou\sthanks{} &
\centering Herv\'e Glotin\sthanks{Supported by GDR CNRS MADICS, bioacoustics group, and http://sabiod.org}
\end{tabular}
}
\address{
\begin{tabular}{p{4.5cm}p{4.5cm}p{4.1cm}p{4.1cm}}
\centering Machine Listening Lab,\newline Centre for Digital Music,\newline Queen Mary University of London. &
\centering Ecosystems and Environment Research Centre,\newline School of Environment and Life Sciences,\newline University of Salford. &
\centering Computer Science Department,\newline University of Crete. &
\centering LSIS UMR CNRS, \newline University of Toulon, \newline Inst. Univ. de France
\end{tabular}
}

\begin{document}

\maketitle

\abstract{
Many biological monitoring projects rely on acoustic detection of birds.
Despite increasingly large datasets, this detection is often manual or semi-automatic,
requiring manual tuning/postprocessing.
We review the state of the art in automatic bird sound detection,
and identify a widespread need for tuning-free and species-agnostic approaches.
We introduce new datasets and an IEEE research challenge
to address this need, to make possible the development of fully automatic algorithms for bird sound detection.
}

\section{Introduction}

Monitoring birds by their sound is important for many environmental and scientific purposes.
A variety of crowdsourcing and remote-monitoring projects now record these sounds,
and some analyse the sound automatically---yet there are still many issues to solve,
as indicated by the number of projects that are yet to be fully automated \cite{Marques:2012,Buxton:2012,Digby:2013}.
In this paper we review research on the specific topic of bird \textit{detection} in audio.
The audio modality is well-suited to bird monitoring because many birds are much more clearly detectable by sound than by vision or other indicators.
We overview the paradigms and techniques used for bird audio detection,
and specific issues to be addressed.
We then describe a data challenge which we are introducing,
with new public datasets, as an initiative to advance the state of the art.
First, though, we must outline the applications for which bird detection in audio is useful.

Bioacoustics has in recent years become one of the ``big data'' research areas,
in particular with remote acoustic monitoring projects generating terabytes of audio,
far more than can feasibly be inspected manually.
The goal of such projects is usually to monitor population densities and migration patterns of animal species,
or to monitor overall ecosystem health.
For example \cite{Borker:2014} found that automatically-detected calling activity was a reliable indicator of relative abundance for monitoring a seabird colony.
\cite{Marques:2012} further reviewed the use of passive acoustic monitoring to estimate animal population density and there is increasing interest in using acoustic indices for biodiversity assessments \cite{Sueur:2014}.

Other large-scale monitoring programmes have used an \textit{occupancy} framework, meaning that instead of working with abundance data (i.e.\ estimated numbers of individuals),
the simpler presence/absence of a species in a spatio-temporal window is the basic observation \cite{Furnas:2015,Rowe:2015ibac}.
The efficiency of collecting occupancy data motivates its use in large-scale studies \cite{Furnas:2015}.
Rowe \cite{Rowe:2015ibac}, using an occupancy framework, found that automated recognition software improves detectability for a range of bird species' vocalizations, though also found that with current technology the manual effort required---to set parameters and to check and post-process the results---means that the efficiency in terms of person-hours was actually not reduced relative to a manual survey. This demonstrates that automatic detection is useful in practice but the automation of this requires further development.
Marques et al.\ \cite{Marques:2012} likewise concluded that improvements in automatic detection (and classification) would be desirable, especially with respect to calibration and full automation.

Unattended monitoring is not the only application to require bird detection.
Another common use case is as a pre-filtering step before other automatic analyses such as bird species classification \cite{Graciarena:2011,Stowell:2014b}.
It is particularly needed in uncontrolled data collection scenarios such as crowdsourcing.

\if\longversion1
Big data problems also affect \textit{manual} work with audio data,
such as the manual browsing and data-mining of audio archives%
\if\longversion1 \cite{Ranft:2004}\fi.
In these situations, content may be sparsely represented,
and the exact audio signal of interest may vary according to the user's query.
Tools are therefore needed to help a person to navigate quickly through large audio collections%
\if\longversion1 \cite{Towsey:2014}\fi.
\fi

These diverse use cases have broadly similar requirements,
but can differ in the exact precision of detail required.
Hence, before reviewing technical approaches used to address these tasks,
we must be a little more specific about the task specifications.

\section{Task paradigms}

\iftrue 
	\begin{figure*}[t] 
		\begin{center}
		\includegraphics [width=0.99\textwidth,clip, trim=20mm 40mm 0mm 18mm]{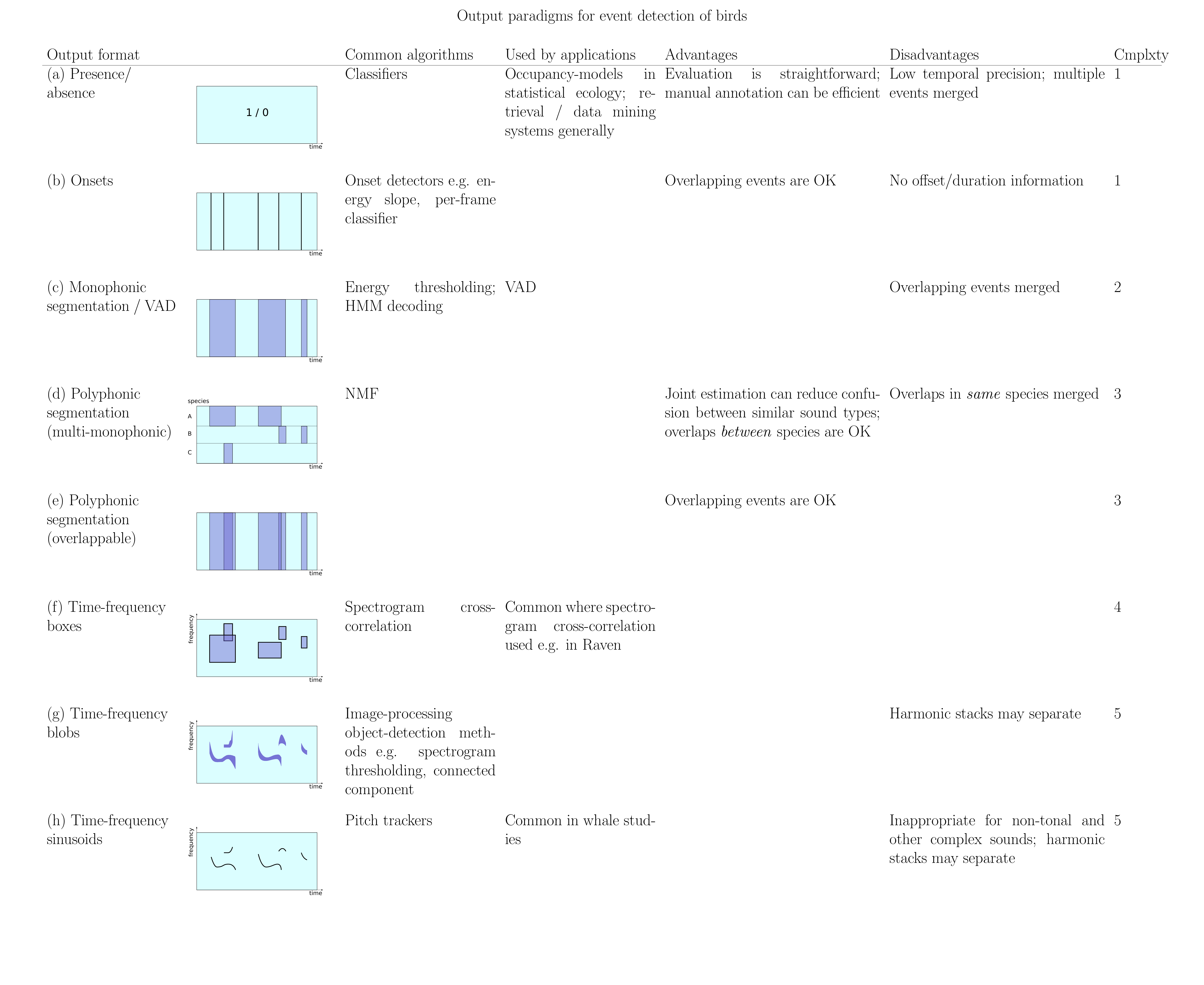}
		\end{center}
		\caption{Task paradigms for bird detection. The final column gives a rough ordering in ascending complexity/difficulty.}
		\label{fig:paradigms}
	\end{figure*}
\else
	\begin{figure}[t]
		\begin{center}
		\includegraphics [width=0.9\linewidth,clip, trim=30mm 40mm 240mm 18mm]{fig_bdetect_paradigms}
		\end{center}
		\caption{Task paradigms for bird detection. The final column gives a rough ordering in ascending complexity/difficulty.}
		\label{fig:paradigms}
	\end{figure}
\fi

The idea of detecting birds in audio can be made into a concrete task in various ways,
each connecting with different application tasks and implying very different output data.
Do we need to know the exact start/end times?
Do we need to know about each vocalisation separately?
Do we need to know how many vocalisations, or how many individuals, or just an overall presence/absence?
Figure \ref{fig:paradigms} illustrates different task paradigms that have been studied, along with some of their characteristics.
%
%
%
%
%
%
%
Their differing characteristics have strong implications,
both for which computational approaches are appropriate and for the practical feasibility of annotation and evaluation.

The most basic is the simple estimation of presence/absence in a given sound clip:
a detector outputs a zero if none of the target species are detected, and a one otherwise.
This provides relatively little information---low temporal detail, no differentiation between one and many detections---%
yet it has practical relevance.
The \textit{occupancy modelling} framework in statistical ecology \cite{Mackenzie:2006} uses exactly this type of data,
and is recommended because it is often easier and less expensive to collect than abundance data,
especially in the context of very large surveys \cite{Furnas:2015,Rowe:2015ibac}.
The output format is a simple binary decision,
which gives scope for various classification methods to be adapted directly,
and also allows for very efficient manual annotation.
For applications such as filtering a large dataset,
or assisting with manual data browsing,
annotated fine detail is often unnecessary since the purpose is simply to help a person or a machine skip over the (typically large) number of negative instances to focus on the audio region containing the positive instances.

A common variant is to add temporal detail to the presence/absence decision:
in other words, to partition the time axis into positive and negative regions (Figure \ref{fig:paradigms} c).
This is often the format produced by methods such as thresholding based on short-term energy levels,
and is appealing for streaming applications such as real-time detectors.
It is analogous to the approach commonly adopted in voice activity detection (VAD) for speech \cite{Ramirez:2007,Ferroni:2015,Zhang:2016}.
Note that in VAD applications there is typically one dominant source of interest,
while in natural sound monitoring the signals of interest are often intrinsically polyphonic.
Thus a positive region may contain one or many vocalisations together.
This task paradigm maintains the advantage of relative simplicity while adding a little more temporal resolution.

Related methods deal with polyphony more explicitly.
Template-matching methods (see next section) yield either event detections (Figure \ref{fig:paradigms} b) or time-frequency boxes (Figure \ref{fig:paradigms} f),
allowing overlapping vocalisations to be represented as separate data points.
Occasionally other methods output temporal regions which can overlap \cite{Stowell:2015b}.
\if\longversion1
Some approaches are polyphonic in a way which we might describe as ``multi-monophonic'' (Figure \ref{fig:paradigms} d):
methods such as non-negative matrix factorisation (NMF) are designed around having multiple source types each of which can be active/inactive,
yet within each source type they do not segregate signals that may overlap or come from multiple individuals
\cite{Benetos:2016}.
These approaches have not received much attention in research on animal sound detection.
\fi

In bird sound, a paradigm that has become common and built in to standard software is to describe events via time-frequency boxes (Figure \ref{fig:paradigms} f).
These can be annotated relatively intuitively by drawing boxes on a spectrogram,
and detected using template-matching methods.
This paradigm works well when sounds are compact in time and frequency:
the approach is not seen in speech and music analysis,
because in those cases the signals of interest are often broadband,
consisting of harmonic stacks and noises.
A sizeable portion of bird sounds is relatively bandlimited;
however for sounds with significant energy across a range of harmonics,
there may be a tendency to exclude higher harmonics,
or to create large regions containing many subbands with no energy from the signal of interest.
These are not show-stopping issues, but may inhibit accuracy for some species.

Going to even more detail, some authors consider detections as arbitrarily-shaped regions on a spectrogram (Figure \ref{fig:paradigms} g)
\cite{Neal:2011,Towsey:2012}.
This approach fits with object-detection methods in image processing.
Often each detected event is required to be a single fully-connected region (``blob detection''),
which is problematic for harmonic sounds since harmonics may then be detected as separate events.
Manually labelling data at this resolution is labour-intensive,
and it is rare that the final downstream applications require such detail.

Tonal sounds can be considered as time-varying sinusoids,
in which case annotation may come in the form of frequency tracks (Figure \ref{fig:paradigms} h).
This has been explored in marine mammal detection,
and sometimes for bird sounds \cite{Jancovic:2011b}.
Again, harmonics might be detected as separate events;
however some models are able to conjoin harmonics into unified tracks.
\if\longversion1
Also note that a sizable number of bird sounds are non-tonal,
and the very rapid pitch modulation of some bird sounds can cause problems for standard frequency trackers.
(The tracking methods of \cite{Jancovic:2011b,Bonada:2016} account for this issue.)
\fi

The task paradigms just discussed each have different affordances.
They provide varying levels of detail for downstream tasks,
but they also enable different sets of technical solutions,
and imply different amounts of manual labour to annotate and evaluate.
We will return to the relative importance of these task paradigms
after reviewing the literature on technical methods that have been used to address them.

\section{Technical methods}

\subsection{Established/baseline methods}

The most common methods for detection are based on either energy, spectrogram cross-correlation, or hidden Markov models (HMMs).
These well-known baselines are available in widely-used bioacoustics software (\textit{Raven}, \textit{XBAT}, \textit{Song Scope}),
and have been used for various surveys.

Perhaps the simplest method is energy thresholding, which yields a VAD-like segmentation output:
positive if the energy in a short time-window is higher than a threshold, otherwise negative.
For bioacoustic surveys it is usually preceded by some kind of noise reduction,
and often applied to bandlimited frequency regions of interest \cite{Duan:2013,McIlraith:1997}.
\cite{Fagerlund:2007} augment it with an iterative process which estimates the background noise level as it converges.

Also common is spectrogram cross-correlation, an alternative which uses one or more example sounds as templates.
These templates are species-specific, and used to scan a spectrogram for regions with strongly-matching profiles by cross-correlation.
For example \cite{Borker:2014} used spectrogram cross-correlation (in XBAT) to detect calls in a seabird colony.
Across millions of calls the detection accuracy was 53.6\%, which the authors compare against 22\%, 17\% and 24\% for studies detecting terrestrial birds; the authors also reported that soundscape characteristics had an impact on detection.

Hidden Markov models (HMMs) have widely been used for sound sequence analysis,
especially in speech,
and have a particular appeal of temporal flexibility that goes beyond template matching.
HMMs have been used for various purposes in bioacoustics including bird detection,
for example in \textit{Song Scope} software \cite{Duan:2013}.
Using \textit{Song Scope}, \cite{Buxton:2012} reported sensitivity ranging from 56\% to 69\% for detecting three seabird species.
\cite{Towsey:2012} report in passing that they found MFCCs and HMMs to perform poorly for detection,
hence their use of other methods.

Alternatively, some investigators use template matching, but instead of cross-correlation they employ dynamic time warping (DTW) which allows for the template and the target sound to be slightly warped in time relative to each other \cite{Anderson:1996}.
DTW is in principle a very suitable method for natural sounds such as bird sounds which often have much organic variability.
However, note that DTW remains much less widely used than cross-correlation, perhaps because in practice the flexibility does not give a strong enough boost over the simpler (and thus computationally more efficient) cross-correlation.

\if\longversion1
For pitched sounds, an alternative detection criterion could be the degree of ``pitch clarity'' in the signal (e.g.\ \cite{Ranjard:2008}, \cite{Lakshminarayanan:2009}).
Such measures are likely to be vulnerable to the effects of noise, which often affect the extent to which the tonal component stands out from the background.
\fi

\subsection{Recent work}

Towsey et al.\ \cite{Towsey:2012} provide a software toolbox with multiple detection methods, including:
spectrogram template matching;
``oscillation detection'' by detecting amplitude modulations in narrow frequency bands; 
energy-based segmentation;
spectral peak tracking;
and spectrogram blob detection.
Note that the focus of Towsey et al.\ is explicitly on single-species targeted studies,
and the choice of method must be chosen based on knowledge of the target species vocalisations.
This is eminently possible for targeted single-species studies,
but difficult to generalise to species-agnostic detection.
It is an interesting open question whether the various different approaches can simply be aggregated under a meta-algorithm to produce species-agnostic output; to our knowledge this has not been attempted.

\cite{Jancovic:2011,Jancovic:2011b} 
detect sinusoidal (pure-tone-like) signals in noise, and use these as a basis for detecting bird species. Their method is able to detect very fast-modulated pitch tracks, which sets it apart from other methods and makes it suitable for many frequency-modulated bird sounds. However, a method based on sinusoidal tracks may of course be inappropriate for the case of non-tonal bird sounds.

\cite{Stowell:2015b} 
propose a model based on detecting onsets and offsets separately, then using typical syllable durations to unify the onsets and offsets into probabilistically smoothed event detections, which may include overlaps. This method is perhaps most suitable for monosyllabic vocalisations.

\if\longversion1
\cite{Papadopoulos:2015} propose to detect presence/absence by training a Gaussian mixture model (GMM) for each chosen species applied to spectral features (such as mean, skewness, flatness), and then using a high GMM likelihood as an indicator of presence.
They find the frequency peak a useful feature, with per-species variability in performance.
\fi

\cite{Ross:2013} use a random forest (RF) classifier to make detection decisions for the presence/absence of flight calls. Sound events are first detected with simple bandlimited energy detection, and then the RF method refines these initial decisions by discarding many of the false positives.
\cite{Ross:2013} argue that the random forest algorithm is appropriate for the detection task, because of various properties including its ability to handle polymorphic categories (i.e.\ the positive events do not all have to be of the same type).

\cite{Neal:2011} work within the relatively uncommon paradigm of detecting exactly which pixels in a spectrogram should be labelled as belonging to bird sound. They train a RF classifier to make the pixel-level decisions. This approach has the clear advantages and drawbacks of the paradigm: the system is able to output detailed estimates (spectrotemporal shapes), at the cost that the user must train the system by providing a set of pixel-wise binary mask information. However, they demonstrate that in this paradigm, the RF achieves much better results than energy-based detection.
\cite{Towsey:2012} also include a pixel-wise method in their toolbox,
based on energy and size of `blob'
rather than on training a classifier,
which should thus be more general though potentially less accurate.
\cite{deOliveira:2015} also use a smoothing technique which sits with these image-based approaches, in their case to preprocess the spectrogram before applying energy-based segmentation. 


There are of course detection systems developed and deployed for related tasks outside of bird sound, in speech, music and environmental sound.
In speech, VAD is well-studied and should be a source of inspiration,
although some methods may be speech-specific
\cite{Ramirez:2007,Ferroni:2015,Zhang:2016}. 
Note also that VAD is typically applied in a monophonic close-mic scenario,
whereas we often wish to detect polyphonic and distant bird sounds.
\cite{Graciarena:2011} use ``a simple voice activity detection system, with acoustic models trained with bird vocalization data'' as a preprocessing step before bird species classification. However the VAD method is not specified.

For general soundscapes,
\cite{Stowell:2015} 
review the state of the art in detecting everyday sound events in urban sound scenes,
and evaluate many methods via a public data challenge using audio recorded in office environments.
The challenge uses two detection paradigms: monophonic and ``multi-monophonic'' (in our current terminology),
in both cases aiming to retrieve the start time, end time and label for each event.
The generally best-performing detector in their study was a two-layer HMM approach;
also strong was a combined HMM and RF method.
MFCCs were not found to be useful features for event detection, generally outperformed by spectrogram or filterbank features.
Regarding evaluation, the authors conclude that more work is needed to ensure that the evaluation measures used for structured data tasks match up with the task desiderata.

\if\longversion1
More recently, \cite{Benetos:2016} develop a system for detecting multiple event types through a probabilistic model with HMM-like constraints on each ``channel''. The model has been used both for music and urban soundscapes.  
\fi


\section{Practical considerations} 

\textbf{Noise reduction; weather noise:}
Background noise must be a consideration,
and even simple noise reduction can help with downstream processing.
The assumption of temporally constant background noise levels (see e.g.\ \cite{Fagerlund:2007})
is in general unrealistic for outdoor sound recordings,
but is a first step for simple noise reduction.
Some approaches allow for smoothly-varying background noise.
However, the bigger issue is robustness to strongly-varying noise,
especially wind and rain,
but also from other fauna (e.g.\ \cite{Digby:2013,Buxton:2012}).
In practical applications heavily affected sound clips may have to be removed (e.g.\ 2.9\% of recording time in \cite{Buxton:2012}).
For example the toolbox of \cite{Towsey:2012} performs automatic wind and rain detection as a classification task.

\textbf{Manual intervention: calibration, post-processing:}
An important issue for large-scale studies and for general application is how much user intervention is actually needed in practice,
even for nominally automatic methods.
Widely-used tools such as Raven and SongScope require manual calibration of thresholds and/or templates for each species of interest before they can be used,
and this can have a strong effect on precision and sensitivity \cite{Duan:2013,Rowe:2015ibac,Buxton:2012}.

\textbf{Single-species vs.\ generic:}
An important question is whether the detector for a certain situation should be detecting individuals from a single species,
or more generally such as from a whole taxon.
Single-species studies can be useful for example in studying so-called ``keystone species'',
or in cases where highly custom detectors might be used to detect idiomatic sounds (e.g.\ woodpecker drumming).
Conversely, there are many cases in which the desire is to detect all vocalisations irrespective of species:
e.g.\ for overall ecosystem monitoring, or as a filtering front-end before further analysis such as classification.
This is particularly the case in situations where not all species are known or well-characterised.

Some techniques are inherently more suited to one or the other:
template detection is inherently specific,
while energy-based detection can be very generic.
For surveys that must cover a wide range of species yet with high specificity
(and perhaps with a high rejection of distractor events),
it may be useful to apply a range of focussed detectors and then to aggregate their outputs.
This can be done straightforwardly.
There is unexplored scope however for meta-algorithms to aggregate the outputs of multiple detectors intelligently,
helping to mitigate ``double firing'' from independent detectors.

\if\longversion1
\section{Evaluations of bird detection}

Surveying studies have evaluated the detection performance of humans \cite{Furnas:2015,Johnston:2014} and more recently of automatic methods.


\cite{Rowe:2015ibac}, using an occupancy framework, found that automated recognition software improves detectability for a range of bird species' vocalizations.
However they also found that with current technology the manual effort required---to set parameters and to check and post-process the results---means that the efficiency in terms of person-hours is actually not reduced. This demonstrates that automatic detection is useful in practice but the need for automation is not yet solved.
\cite{Borker:2014} also reported that financial costs were approximately equal for manual and automated nest census monitoring,
though in that case they highlighted the initial equipment setup costs as the main issue.

\cite{Digby:2013} compared manual and autonomous acoustic detection for kiwis. They found autonomous detection could be highly reliable and very efficient in person-time. They noted that automatic recorders were much more strongly affected by weather (especially wind) than were humans.
\cite{Hutto:2009} conducted a multi-species comparison of in-field vs.\ remote methods for point-count surveys. Remote monitoring was performed by autonomous recording units, and the audio recordings were manually labelled by a human observer through audio and spectrograms inspected using Raven. They found that this method performed much worse than the human observers, suggesting that it was not a cost-effective means of gathering survey data.
However \cite{Furnas:2015} found that methods based on automated sound recording have the sensitivity necessary e.g.\ for detecting population decline to a required standard.

\fi

\section{A research data challenge}

To stimulate the next research advances on species-agnostic bird detection,
we present an IEEE-sponsored data challenge.
For this challenge we introduce two new public datasets of annotated audio data.
%
%
For the challenge tasks we have opted for the presence/absence paradigm, applied to ten-second audio excerpts.
As discussed, this approach fits well with statistical applications such as the occupancy framework,
is efficient for manual annotation,
and has clear evaluation.
(cf. \cite{Towsey:2012} using the same paradigm.)
It can be addressed by a wide variety of approaches.

\subsection{Datasets}

Our first dataset comes from a UK bird-sound crowdsourcing research spinout called Warblr.%
\footnote{\url{http://warblr.net}}
From this initiative we have over 10,000 ten-second smartphone audio recordings
from around the UK. The audio totals around 28 hours duration. The audio will be published by
Warblr under a Creative Commons licence. The audio covers a wide distribution of UK locations
and environments, and includes weather noise, traffic noise, human speech and even human bird
imitations. It is directly representative of the data that is collected from a mobile crowdsourcing
initiative.
Annotations of the Warblr dataset are performed by a network of volunteers.

Our second dataset comes from the TREE (Transfer-Exposure-Effects) research project (www.ceh.ac.uk/TREE), 
which is funded by the Natural Environment Research Council (NERC), Environment Agency and Radioactive Waste Management Ltd.  
Dr Mike Wood's team are using unattended acoustic recorders in the Chernobyl Exclusion Zone (CEZ) to capture the Chernobyl soundscape and investigate 
the long-term effects of the Chernobyl accident on the local ecology.  To date, the study has captured approximately 10,000 hours of audio from the CEZ.  Dr Wood's 
team are annotating a data subset for bird species presence/absence, and approx 48---72 hours of annotated audio will be
made available for the BAD Challenge. The audio covers a range of birds and includes weather,
large mammal and insect noise sampled across various CEZ environments, including abandoned
village, grassland and forest areas.

To provide an initial indication of the general level of difficulty within a single dataset, we ran a two-fold
crossvalidation test using a subset of the Warblr data and a simple baseline binary classifier.
We used the baseline previously created for the DCASE challenge, a generic MFCC+GMM pipeline as used for various audio tasks in the past \cite{Stowell:2015}.
In a previous study, this baseline system achieved 82\% AUC in an auto-tagging study to detect the ``birdsong'' tag in audio soundscapes \cite{Stowell:2014f}.
In the present case, the baseline attained a similar value of 79\% AUC---above the 50\% chance level but with substantial headroom for the challenge.
Recall that this test is to detect presence/absence across potentially hundreds of bird species,
making it rather impractical to use certain single-species methods.

\subsection{Organisation}

As is typical for data challenges, we will partition the data and annotations into training, validation and testing partitions, with the testing annotations kept private for evaluation.
Further, we will incentivise the development of generalisable and ``tuning-free'' methods by
ensuring that at least one set of testing data is recorded under different conditions than the
publicly-available data.
This will create a harder task than the within-dataset task for which the AUCs above were measured (further baselines, for these cases, will be published later).
This helps ensure that the challenge addresses the need for methods that work with minimal manual intervention, as identified in the review we present here.

Participants will be challenged to create a system that can label the presence/absence across a diverse species range; they will not be required to identify the species.
The data will be released in Summer 2016,
with a deadline of late 2016 for challenge submissions.
Results will be presented at a conference special session in 2017.
For more detail on the timeline, please visit the challenge website.%
\footnote{\url{http://machine-listening.eecs.qmul.ac.uk/bird-audio-detection-challenge/}}

\section{Conclusions}

This survey has described current approaches to automatic bird detection in audio,
including the current level of generality.
Open topics include weather robustness and tuning-free methods.
We have introduced a challenge giving researchers an opportunity to create a step change in these directions.
A wide variety of methodological options remains open to further study,
such as recent innovations in deep learning%
\if\longversion1
{} \cite{Lecun:2015}%
\fi,
or meta-algorithms that can automatically select detectors
or combine their outputs.

\bibliographystyle{IEEEbib}
\bibliography{bdetect_trimmed}

\begin{thebibliography}{10}

\bibitem{Marques:2012}
T.~A. Marques et~al.,
\newblock ``Estimating animal population density using passive acoustics,''
\newblock {\em Biol Reviews}, 2012.

\bibitem{Buxton:2012}
R.~T. Buxton and I.~L. Jones,
\newblock ``Measuring nocturnal seabird activity and status using acoustic
  recording devices: applications for island restoration,''
\newblock {\em J Field Ornithology}, vol. 83, no. 1, pp. 47--60, 2012.

\bibitem{Digby:2013}
A.~Digby et~al.,
\newblock ``A practical comparison of manual and autonomous methods for
  acoustic monitoring,''
\newblock {\em Meth Ecol Evol}, vol. 4, no. 7, pp. 675--683, 2013.

\bibitem{Borker:2014}
A.~L. Borker et~al.,
\newblock ``Vocal activity as a low cost and scalable index of seabird colony
  size,''
\newblock {\em Conservation Biology}, 2014.

\bibitem{Sueur:2014}
J.~Sueur, A.~Farina, A.~Gasc, N.~Pieretti, and S.~Pavoine,
\newblock ``Acoustic indices for biodiversity assessment and landscape
  investigation,''
\newblock {\em Acta Acustica united with Acustica}, vol. 100, no. 4, pp.
  772--781, 2014.

\bibitem{Furnas:2015}
B.~J. Furnas and R.~L Callas,
\newblock ``Using automated recorders and occupancy models to monitor common
  forest birds across a large geographic region,''
\newblock {\em J Wildlife Management}, vol. 79, no. 2, pp. 325--337, 2015.

\bibitem{Rowe:2015ibac}
K.~M.~C. Rowe,
\newblock ``Automated recognition software improves detectability for a range
  of bird species' vocalizations,''
\newblock in {\em Int Bioacoustics Congress (IBAC)}, 2015.

\bibitem{Graciarena:2011}
M.~Graciarena et~al.,
\newblock ``Bird species recognition combining acoustic and sequence
  modeling,''
\newblock in {\em Proc ICASSP}, 2011, p. 341.

\bibitem{Stowell:2014b}
D.~Stowell and M.~D. Plumbley,
\newblock ``Automatic large-scale classification of bird sounds is strongly
  improved by unsupervised feature learning,''
\newblock {\em PeerJ}, vol. 2, pp. e488, 2014.

\bibitem{Ranft:2004}
R.~Ranft,
\newblock ``Natural sound archives: Past, present and future,''
\newblock {\em Anais da Academia Brasileira de Ci{\^e}ncias}, vol. 76, no. 2,
  pp. 456--460, 2004.

\bibitem{Towsey:2014}
M.~Towsey et~al.,
\newblock ``Visualization of long-duration acoustic recordings of the
  environment,''
\newblock {\em Procedia Computer Science}, vol. 29, pp. 703--712, 2014.

\bibitem{Mackenzie:2006}
D.~I. MacKenzie et~al.,
\newblock {\em Occupancy estimation and modeling: inferring patterns and
  dynamics of species occurrence},
\newblock Elsevier/Academic Press, Burlington, MA, 2006.

\bibitem{Ramirez:2007}
J.~Ram{\'\i}rez, JM~G{\'o}rriz, and JC~Segura,
\newblock ``Voice activity detection. fundamentals and speech recognition
  system robustness,''
\newblock in {\em Robust Speech Recognition and Understanding}, M.~Grimm and
  K.~Kroschel, Eds., chapter~1. 2007.

\bibitem{Ferroni:2015}
G.~Ferroni, R.~Bonfigli, E.~Principi, S.~Squartini, and F.~Piazza,
\newblock ``A deep neural network approach for voice activity detection in
  multi-room domestic scenarios,''
\newblock in {\em 2015 Int Joint Conf on Neural Networks (IJCNN)}. IEEE, 2015,
  pp. 1--8.

\bibitem{Zhang:2016}
X.-L. Zhang and D.~Wang,
\newblock ``Boosting contextual information for deep neural network based voice
  activity detection,''
\newblock {\em IEEE/ACM Trans Audio, Speech, and Language Processing}, vol. 24,
  no. 2, pp. 252--264, 2016.

\bibitem{Stowell:2015b}
D.~Stowell and D.~Clayton,
\newblock ``Acoustic event detection for multiple overlapping similar
  sources,''
\newblock in {\em Proc WASPAA 2015}, 2015.

\bibitem{Benetos:2016}
E.~Benetos et~al.,
\newblock ``Detection of overlapping acoustic events using a
  temporally-constrained probabilistic model,''
\newblock in {\em ICASSP}, 2016.

\bibitem{Neal:2011}
L.~Neal et~al.,
\newblock ``Time-frequency segmentation of bird song in noisy acoustic
  environments,''
\newblock in {\em Proc ICASSP}, 2011, pp. 2012--2015.

\bibitem{Towsey:2012}
M.~Towsey et~al.,
\newblock ``A toolbox for animal call recognition,''
\newblock {\em Bioacoustics}, vol. 21, no. 2, pp. 107--125, 2012.

\bibitem{Jancovic:2011b}
P.~Jan{\v{c}}ovi{\v{c}} and M.~K{\"o}k{\"u}er,
\newblock ``Automatic detection and recognition of tonal bird sounds in noisy
  environments,''
\newblock {\em EURASIP J Advances in Sig Proc}, vol. 2011, no. 1, pp. 982936,
  2011.

\bibitem{Bonada:2016}
J.~Bonada, R.~Lachlan, and M.~Blaauw,
\newblock ``Bird song synthesis based on hidden markov models,''
\newblock in {\em Proceedings of InterSpeech 2016}, 2016.

\bibitem{Duan:2013}
S.~Duan et~al.,
\newblock ``Timed probabilistic automaton: a bridge between raven and song
  scope for automatic species recognition,''
\newblock in {\em Proc. 25th Innovative Applications of Artificial Intelligence
  Conf}, 2013, pp. 1519--1524.

\bibitem{McIlraith:1997}
A.~L. McIlraith and H.~C. Card,
\newblock ``Birdsong recognition using backpropagation and multivariate
  statistics,''
\newblock {\em IEEE Trans Sig Proc}, vol. 45, no. 11, pp. 2740--2748, 1997.

\bibitem{Fagerlund:2007}
S.~Fagerlund,
\newblock ``Bird species recognition using support vector machines,''
\newblock {\em EURASIP J Applied Sig Proc}, p. 38637, 2007.

\bibitem{Anderson:1996}
S.~E. Anderson, A.~S. Dave, and D.~Margoliash,
\newblock ``Template-based automatic recognition of birdsong syllables from
  continuous recordings,''
\newblock {\em J Acoustical Soc America}, vol. 100, no. 2, Part 1, pp.
  1209--1219, 1996.

\bibitem{Ranjard:2008}
L.~Ranjard and H.~A. Ross,
\newblock ``Unsupervised bird song syllable classification using evolving
  neural networks,''
\newblock {\em Journal of the Acoustical Society of America}, vol. 123, no. 6,
  pp. 4358--4368, Jun 2008.

\bibitem{Lakshminarayanan:2009}
B.~Lakshminarayanan, R.~Raich, and X.~Fern,
\newblock ``A syllable-level probabilistic framework for bird species
  identification,''
\newblock in {\em Proceedings of the 2009 International Conference on Machine
  Learning and Applications}, 2009, pp. 53--59.

\bibitem{Jancovic:2011}
P.~Jan{\v{c}}ovi{\v{c}} and M.~K{\"o}k{\"u}er,
\newblock ``Detection of sinusoidal signals in noise by probabilistic modelling
  of the spectral magnitude shape and phase continuity,''
\newblock in {\em Proc ICASSP}, 2011, pp. 517--520.

\bibitem{Papadopoulos:2015}
T.~{Papadopoulos}, S.~{Roberts}, and K.~{Willis},
\newblock ``{Detecting bird sound in unknown acoustic background using
  crowdsourced training data},''
\newblock {\em ArXiv e-prints}, May 2015.

\bibitem{Ross:2013}
J.~C. Ross and P.~E. Allen,
\newblock ``Random forest for improved analysis efficiency in passive acoustic
  monitoring,''
\newblock {\em Ecological Informatics}, 2013.

\bibitem{deOliveira:2015}
A.~G. de~Oliveira et~al.,
\newblock ``Bird acoustic activity detection based on morphological filtering
  of the spectrogram,''
\newblock {\em Applied Acoustics}, vol. 98, pp. 34--42, 2015.

\bibitem{Stowell:2015}
D.~Stowell et~al.,
\newblock ``Detection and classification of acoustic scenes and events,''
\newblock {\em {IEEE} Trans Multimedia}, vol. 17, no. 10, pp. 1733--1746,
  October 2015.

\bibitem{Johnston:2014}
A.~Johnston et~al.,
\newblock ``Species traits explain variation in detectability of uk birds,''
\newblock {\em Bird Study}, vol. 61, no. 3, pp. 340--350, 2014.

\bibitem{Hutto:2009}
R.~L. Hutto and R.~J. Stutzman,
\newblock ``Humans versus autonomous recording units: a comparison of
  point-count results,''
\newblock {\em J Field Ornithology}, vol. 80, no. 4, pp. 387--398, 2009.

\bibitem{Stowell:2014f}
D.~Stowell and M.~D. Plumbley,
\newblock ``An open dataset for research on audio field recording archives:
  freefield1010,''
\newblock in {\em Proc AES53}. 2014, Audio Engineering Society.

\bibitem{Lecun:2015}
Y.~LeCun, Y.~Bengio, and G.~Hinton,
\newblock ``Deep learning,''
\newblock {\em Nature}, vol. 521, no. 7553, pp. 436--444, 2015.

\end{thebibliography}
\end{document}